\documentclass[notitlepage,aps,nofootinbib,letterpaper,prx,twocolumn,showpacs,amsmath,amssymb,amsfonts,long,superscriptaddress,preprintnumbers,longbibliography]{revtex4-1}

\usepackage{amsmath,amssymb,amsthm,amsxtra,overpic,bbm,bm,epsfig,subfigure}
\usepackage{simplewick} 
\usepackage[dvipsnames]{xcolor}
\usepackage{epstopdf}
\usepackage{epsfig}
\usepackage{comment}
\usepackage{graphicx,color,soul}
\usepackage{multirow}
\usepackage{verbatim}
\usepackage{enumitem}

\usepackage{bold-extra}
\usepackage[hidelinks]{hyperref}

\begin{document}

\title{Hunting the Glashow Resonance with PeV Neutrino Telescopes}

\author{Guo-yuan Huang}
\email{huanggy@ihep.ac.cn} 
\affiliation{Institute of High Energy Physics, and School of Physical Sciences}
\affiliation{University of Chinese Academy of Sciences, Beijing 100049, China}
\author{Qinrui Liu}
\email{qliu246@wisc.edu    (Corresponding Author)}
\affiliation{Department of Physics and Wisconsin IceCube Particle Astrophysics Center, University of Wisconsin-Madison,
Madison, WI 53706, USA}

\date{\today}

\begin{abstract}
The Glashow resonant scattering, i.e. ${\overline{\nu}^{}_{e} + e^{-} \rightarrow W^{-} \rightarrow \text{anything}}$, offers us a possibility of disentangling $\overline{\nu}^{}_{e}$ from the total astrophysical neutrino fluxes.
Meanwhile, a great number of high-energy neutrino telescopes, with various detection mechanisms, are advancing towards a better understanding of one of the most energetic frontiers of the Universe. In this work, we investigate a connection between through-going muons at IceCube and the Glashow resonance signal through the channel $W^{-} \rightarrow \mu$. We find that for IceCube, muons from $\overline{\nu}^{}_{e}$  can induce a $\sim20\%$ excess of PeV events around the horizontal direction. However, the current statistics of IceCube is not enough to observe such an excess. We also address the novel possibility of $\overline{\nu}^{}_{e}$ detection via $W^{-} \rightarrow \tau$ at telescopes aiming to detect Earth-skimming and mountain-penetrating neutrinos. The subsequent hadronic decay of tau will induce an extensive air shower which can be detected by telescopes with Cherenkov or fluorescence techniques.
Similar to IceCube, it is challenging  to observe the Glashow resonance excess from the Earth-skimming neutrinos. Nevertheless, we find it is promising to observe Glashow resonance events with the mountain as the target.
\vspace*{3em}
\end{abstract}

\preprint{}

\maketitle
\bigskip

\section{Introduction}
The detection of high-energy astrophysical neutrinos at IceCube has initiated a new era of multimessenger astronomy \cite{Aartsen:2013jdh}. Unlike messengers such as cosmic rays and gamma rays, the weakly interacting neutrinos originating from astrophysical sources can traverse cosmological distances freely without being attenuated by cosmic relics or bent by magnetic fields. However, the nature of sources of the observed astrophysical neutrinos are as yet unknown. A feasible way to improve our knowledge of the source class is to pin down the flavor ratios of those astrophysical neutrinos at Earth \cite{Mena:2014sja,Chen:2014gxa,Palomares-Ruiz:2015mka,Aartsen:2015ivb,Palladino:2015zua,Arguelles:2015dca,Bustamante:2015waa,Aartsen:2015knd,Bustamante:2019sdb,Palladino:2019pid,Stachurska:2019srh}.
%, namely $\Phi^{\oplus}_{\nu_e}:\Phi^{\oplus}_{\overline{\nu}_e}:\Phi^{\oplus}_{\nu_{\mu}}:\Phi^{\oplus}_{\overline{\nu}_{\mu}}:\Phi^{\oplus}_{\nu_{\tau}}:\Phi^{\oplus}_{\overline{\nu}_{\tau}}$ with $\Phi^{\oplus}_{\nu_{\alpha}}$ and $\Phi^{\oplus}_{\overline{\nu}_{\alpha}}$ being the fluxes of $\nu_{\alpha}$ (for $\alpha = e, \mu, \tau$) and $\overline{\nu}_{\alpha}$, respectively. 
However, at neutrino energies above $\mathcal{O}(100~{\rm TeV})$ the valence-quark contribution in the neutrino-nucleon deep inelastic scattering (DIS) becomes less important, and the difference between DIS cross sections of neutrinos and antineutrinos diminishes \cite{Gandhi:1995tf,Formaggio:2013kya}. Therefore, neutrino telescopes are blind to the discrimination of neutrinos versus antineutrinos.

This degeneracy must be resolved in order to distinguish different astrophysical neutrino sources, e.g., optical thin sources dominated by $p\gamma$ interactions and by $pp$ interactions. In the $p\gamma$ scenario, the boosted protons will interact with ambient photons, e.g. $p + \gamma \rightarrow \pi^{+} + n$ to produce pions. The subsequent decays of $\pi^{+}$ generate neutrinos with a flavor composition of $\Phi^{\rm S}_{\nu_e}:\Phi^{\rm S}_{\nu_{\mu}}:\Phi^{\rm S}_{\overline{\nu}_{\mu}} = 1:1:1$ at the source, with other unspecified flavor ratios being zero. 
$\pi^{-}$ can also be produced in the $p\gamma$ interactions through the multipion production channel, but this would require a much higher center-of-mass energy, and is usually suppressed. If a strong magnetic field is presented in the source environment, $\mu^{+}$ from the $\pi^{+}$ decay will suffer from rapid energy loss due to the cyclotron radiation, and its contribution to the high-energy neutrino flux will be reduced, with dominantly $\Phi^{\rm S}_{\nu_{\mu}}$ left as a result. 
A characteristic for a $p\gamma$ source is the low number of $\overline{\nu}^{}_{e}$ component yield at the source.
On the other hand, for the $pp$ source both $\pi^{+}$ and $\pi^{-}$ can be copiously produced via $pp$ scattering, resulting in equal fractions of neutrinos and antineutrinos, e.g. with $\Phi^{\rm S}_{\nu_e} + \Phi^{\rm S}_{\overline{\nu}_{e}}:\Phi^{\rm S}_{\nu_{\mu}}+\Phi^{\rm S}_{\overline{\nu}_{\mu}} = 1:2$.
After propagation of a cosmological distance, neutrino fluxes of different flavors at Earth will mix with each other according to the relation
$\Phi^{\oplus}_{\nu_{\beta}} = \sum^{}_{\alpha,i} \Phi^{\rm S}_{\nu_{\alpha}} |U^{}_{\alpha i}|^2 |U^{}_{\beta i}|^2$, where $\{\alpha,\beta\}$ run over $\{e,\mu,\tau\}$, and $i$ runs over the mass indices $\{1,2,3\}$. Here $U$ is the unitary lepton flavor mixing matrix, and the global-fit results can be found in Ref.~\cite{Esteban:2018azc}. A symmetry between the $\mu$ flavor and $\tau$ flavor can be noticed, i.e., $|U^{}_{\mu i}| \simeq |U^{}_{\tau i}|$ \cite{Xing:2015fdg}. This always leads to $\Phi^{\oplus}_{\nu_{\mu}} \simeq \Phi^{\oplus}_{\nu_{\tau}}$ and $\Phi^{\oplus}_{\overline{\nu}_{\mu}} \simeq \Phi^{\oplus}_{\overline{\nu}_{\tau}}$ no matter what flavor compositions we start from \cite{Xing:2007zza}. Using the best-fit oscillation parameters, we can obtain the flavor compositions at Earth ($\Phi^{\oplus}_{\nu_e}:\Phi^{\oplus}_{\overline{\nu}_e}:\Phi^{\oplus}_{\nu_{\mu}}:\Phi^{\oplus}_{\overline{\nu}_{\mu}}:\Phi^{\oplus}_{\nu_{\tau}}:\Phi^{\oplus}_{\overline{\nu}_{\tau}}$) for different source models as follows:
(i) $\left\{3.6,\,1,\, 3,\, 2,\, 3,\, 1.8\right\}$ for $p\gamma$ source; (ii) $\left\{1,\, 0,\, 2,\, 0,\, 1.8,\, 0\right\}$ for $\mu$-damped $p\gamma$ source; (iii) $\left\{1,\, 1,\, 1.1,\, 1.1,\, 1,\, 1 \right\}$ for $pp$ source; (iv) $\left\{1,\, 1,\, 2,\, 2,\, 1.8,\, 1.8 \right\}$ for $\mu$-damped $pp$ source.

Another widely expected neutrino flux in the PeV-ZeV energy range is produced by the scattering of cosmic rays with the cosmic photon background, the so-called Greisen-Zatsepin-Kuzmin (GZK) neutrinos \cite{Greisen:1966jv,Zatsepin:1966jv,Beresinsky:1969qj}. At the multi-PeV energies, GZK neutrinos are dominated by the decay of neutrons produced from $p\gamma$ scatterings. Hence, low-energy GZK neutrinos at the source are mainly $\overline{\nu}^{}_{e}$, leading to the flavor composition $\Phi^{\oplus}_{\overline{\nu}_e}:\Phi^{\oplus}_{\overline{\nu}_{\mu}}:\Phi^{\oplus}_{\overline{\nu}_{\tau}} = \left\{ 1,\, 0.4,\, 0.4 \right\}$ at Earth. 

The resonant scattering around the $W$ pole \cite{Glashow:1960zz}, i.e. ${\overline{\nu}^{}_{e} + e^{-} \rightarrow W^{-}_{} \rightarrow \text{hadrons or leptons}}$, provides us a window to discriminate $\overline{\nu}^{}_{e}$ from the total neutrino flux, and therefore distinguish between the $p\gamma$ and $pp$ sources. 
The IceCube observatory at the South Pole has a great potential to measure the $\overline{\nu}^{}_{e}$ flux by observing the event excess around the resonant energy of $6.3~{\rm PeV}$.
The search for astrophysical sources with the Glashow resonance (GR) has already been extensively discussed for IceCube and its future upgrade IceCube-Gen2 \cite{Aartsen:2014njl}, e.g., in Refs.~\cite{Anchordoqui:2004eb,Hummer:2010ai,Xing:2011zm,Bhattacharya:2011qu,Bhattacharya:2012fh,Barger:2012mz,Barger:2014iua,Palladino:2015uoa,Shoemaker:2015qul,Anchordoqui:2016ewn,Kistler:2016ask,Biehl:2016psj}. 
Similar analysis can also be straightforwardly applied to its water counterpart ANTARES \cite{AdrianMartinez:2012rp} in the Mediterranean sea, as well as future KM3NeT \cite{Adrian-Martinez:2016fdl} and Baikal-GVD \cite{Avrorin:2018ijk}.
These previous studies on Glashow resonance mainly focus on the starting events at IceCube, for which the interaction vertex is contained within the detector.
A decent background rejection can be achieved with the good energy reconstruction of cascades at IceCube as well as the identification of some nearly background-free morphology \cite{Bhattacharya:2011qu,Bhattacharya:2012fh}. Interestingly, a partially contained cascade at IceCube may possibly be the first detected Glashow resonance event \cite{Lu:2017nti,lulu}.   

A hunt for the Glashow resonance should be performed using neutrino data in the multi-PeV energy region. The IceCube experiment has reported several events with reconstructed neutrino energy above PeV, including the high-energy starting events (HESE) as well as the through-going muons (TGM). In the publicly available HESE sample \cite{Aartsen:2014gkd,Schneider:2019ayi}, there are three cascade events with deposited energies being $1040.7_{-144.4}^{+131.6}~{\rm TeV}$, $1140.8_{-132.8}^{+142.8}~{\rm TeV}$ and $2003.7_{-261.5}^{ +236.2}~{\rm TeV}$. The primary neutrino energy is slightly higher than the deposited one. However, none of them can be the Glashow resonance event within the  acceptable statistical significance.
On the other hand, the TGM sample contains two unusual track events with the reconstructed muon energies as high as $E^{}_{\mu} \simeq 4.5~{\rm PeV}$ \cite{Aartsen:2016xlq} and $ 1.2~{\rm PeV}$ \cite{Stettner:2019tok}, which should correspond to primary neutrinos with higher energies. 
Unlike the starting events at IceCube, the energy reconstruction is relatively poor for through-going events due to the lack of knowledge of the interacting point outside the detector.
The reconstructed neutrino energy spans a wide posterior range, also subject to the assumptions of the track type \cite{Kistler:2016ask}, the neutrino spectrum and flavor priors. 
In a six-year TGM analysis from 2009 to 2015 in Ref.~\cite{Aartsen:2016xlq},
assuming a best-fit neutrino spectrum,  there should be approximately a dozen TGM events which are induced by multi-PeV neutrinos. Given this channel of the Glashow resonance, e.g. $W^{-} \rightarrow \mu$, it is interesting to ask a question: \textit{are there any through-going events at IceCube that are generated via the Glashow resonance channel?}

Other than the conventional approach of Cherenkov light detection with a large detection volume of ice (IceCube) or water (ANTARES), there are neutrino telescopes with other working principles, sensitive to neutrinos at higher energies \cite{Baret:2011zz}.
A technique is to detect the Askaryan radio emission of in-ice showers \cite{Askaryan:1962hbi,Saltzberg:2000bk,Gorham:2006fy}, which has been used for the neutrino searches by RICE \cite{Kravchenko:2011im}, ANITA \cite{Gorham:2016zah,Allison:2018cxu}, ARA \cite{Allison:2015eky} and ARIANNA \cite{Barwick:2014pca}, and will be adopted in RNO \cite{Aguilar:2019jay}.
Another detection principle relies on the extensive air shower produced by the hadronic decay of tau \cite{Berezinsky:1975zz,Domokos:1997ve,Domokos:1998hz,Capelle:1998zz,Fargion:1999se,Fargion:2000iz,LetessierSelvon:2000kk,Feng:2001ue,Kusenko:2001gj,Bertou:2001vm},
which is usually assumed to be produced by the charged-current interaction of $\nu^{}_{\tau}$ inside Earth. However, \textit{we emphasize that the Glashow resonance is also relevant for tau production at these telescopes, via the channel $W^{-} \rightarrow \tau$}.
This may give us another possible experimental window to the detection of  $\overline{\nu}^{}_{e}$ \cite{Fargion:1999se,Fargion:2000iz}. 
Notice that for through-going track events at IceCube, the interaction vertex is separated from the detection volume, and the IceCube detector in this case plays a very similar role.

There are multiple techniques to detect the air-shower signal from neutrinos, including the detection of particles in the shower \cite{Aab:2015kma,Aab:2019auo},
radio signals \cite{Gorham:2016zah,Alvarez-Muniz:2018bhp,Allison:2018cxu}, fluorescence \cite{Cao:2004sd,Sasaki:2014mwa,Neronov:2019htv} and Cherenkov light \cite{Cao:2004sd,Sasaki:2014mwa,Neronov:2016zou,Ahnen:2018ocv,Krizmanic:2019hiq,Otte:2018uxj,Otte:2019aaf} in the atmosphere. 
The strength of the signal is generically proportional to the initial neutrino energy.
Not all of those techniques can have a multi-PeV neutrino sensitivity, i.e. down to the Glashow resonance.
Among these detection methods, the direct particle detection and the radio detection are sensitive to neutrinos at relatively higher energies, and their sensitivities will decrease significantly for neutrino energies below 10-100 PeV. 
For instance, the energy threshold of the radio arrays of GRAND is limited by the noise level at the antenna output \cite{Alvarez-Muniz:2018bhp}. Without enough primary neutrino energy, the radio-induced voltage may not be enough to exceed the stationary background noise.
In contrast, telescopes using Cherenkov and fluorescence techniques can reach a much lower energy threshold, e.g., down to as low as PeV, without losing too much detection efficiency. Therefore, in this work we shall constrain our analyses to the Cherenkov or fluorescence detection techniques.

For this study, two typical targets for neutrino interactions are considered: (i) the Earth crust near horizon for Earth-skimming neutrinos; (ii) a mountain target for the mountain-penetrating neutrinos.
The remaining part is organized as follows. In Sec.~II, we present the general strategy for this study. The through-going events of IceCube are investigated in Sec.~III. In Sec.~IV and Sec.~V we study the Glashow resonance signatures from Earth-skimming and mountain-penetrating $\overline{\nu}^{}_{e}$, respectively.
Sec.~VI is devoted to a conclusion.

\section{Framework}
\begin{figure}[t!]
	\centering
	\hspace{0cm}
	\includegraphics[width=0.89\columnwidth]{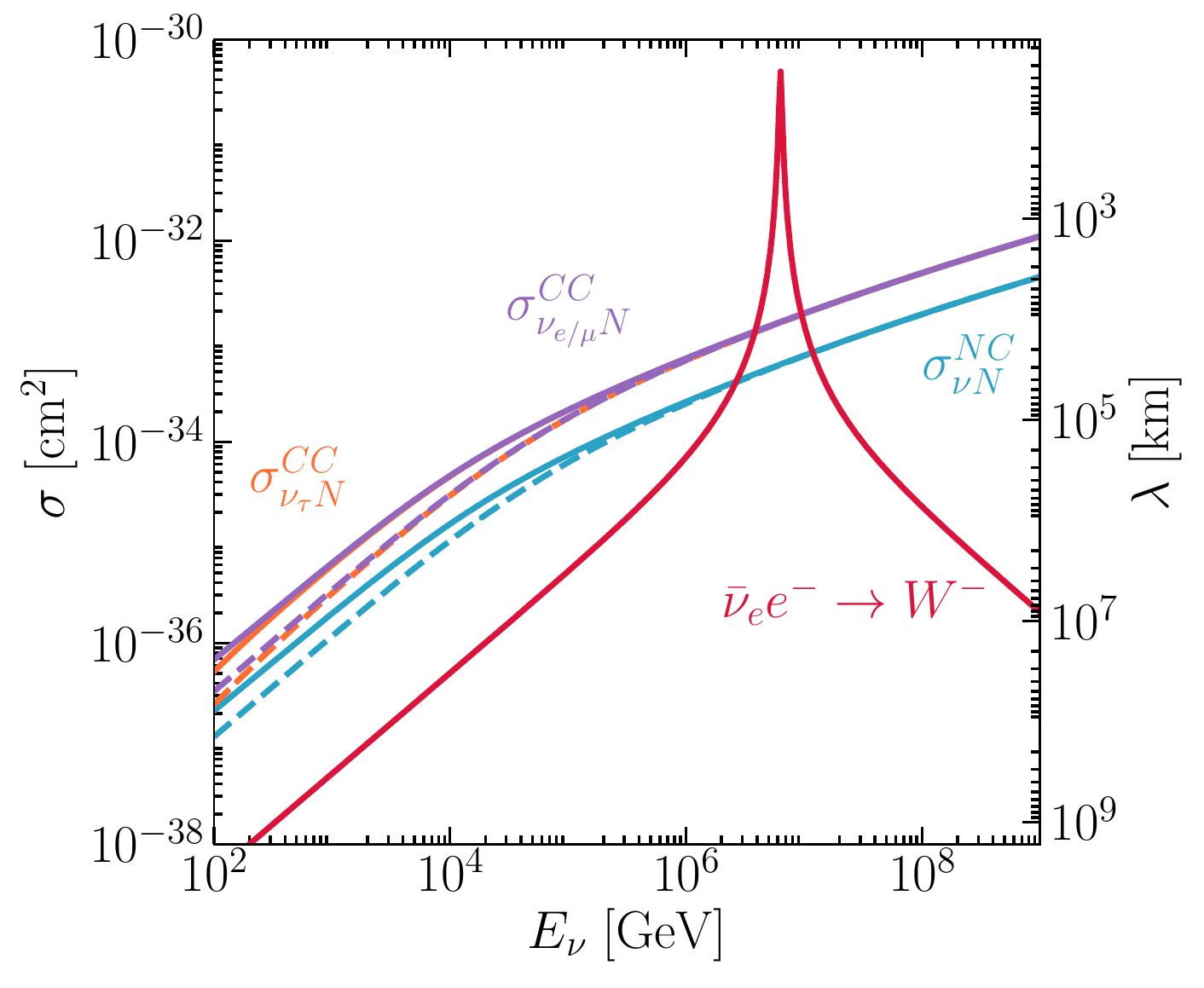}
	\caption{Cross sections of neutrino scattering on a nucleon or electron. The red curve signifies the case of the GR. The deep inelastic scattering processes include the neutral-current interaction for all neutrino flavors (blue curves), and the charged-current interactions for $\nu_\mu/\nu_e$ (purple curves) and for $\nu_\tau$ (orange curves). Solid curves stand for neutrinos and dashed ones for antineutrinos. The y-axis on the right represents the water equivalent MFP of these interactions. \label{fig:xsec}}
\end{figure}

High-energy neutrinos propagating inside Earth undergo collisions with dense matter composed of nucleons and electrons, resulting in a reduction in the flux. 
The attenuation effect becomes significant when the energy of neutrinos is beyond $10~{\rm TeV}$, for which the mean free path (MFP) of neutrinos in matter is comparable to the diameter of the Earth.  With the IceCube data, this effect has already been utilized to set constraints on the neutrino-nucleon scattering cross sections \cite{Hooper:2002yq,
Borriello:2007cs,
Connolly:2011vc,Chen:2013dza,
Klein:2013xoa,Aartsen:2017kpd,Bustamante:2017xuy}. 
There are three dominant interactions within the SM which are relevant when a neutrino traverses the Earth:
(i) the charged-current (CC) DIS with nucleons mediated by the $W$ boson, from which the hadronic shower, as well as an outgoing charged lepton will be produced;
(ii) the neutral-current (NC) DIS mediated by the $Z$ boson, giving rise to a hadronic shower with an energy reduced neutrino emitted;
(iii) the interaction with electrons mediated by $W$ and $Z$ bosons. 
The neutrino-electron cross section in most cases is smaller than the neutrino-nucleon one by orders of magnitude. However, for the scattering of $\overline{\nu}^{}_{e}$ with electrons, when the center-of-mass energy is close to the $W$ pole mass, the interaction can be greatly enhanced via the on-shell production of $W$.
The cross section in this case reads 
\begin{align}
\sigma^{}_{\overline{\nu}^{}_{e}e}(s) 
=\,& 24\pi\,\Gamma^2_W\,{\rm Br}(W^-{\rightarrow}\,\overline{\nu}_e+e^-) \nonumber\\
&\hspace{0cm}\times \frac{s/M^2_W}{(s-M^2_W)^2+(M_W\Gamma_W)^2} \;,
\label{eq:xsec}
\end{align}
where $s$ denotes the square of the center-of-mass energy, $\Gamma^{}_{W}$ is the total decay rate of $W$, and ${\rm Br}(W^-\rightarrow\overline{\nu}_e+e^-) \simeq 10.7\% $ stands for the branch ratio of the decay channel $W^-\rightarrow\overline{\nu}_e+e^-$. The cross section $\sigma^{}_{\overline{\nu}^{}_{e}e}(s)$ reaches its maximum value $\sigma^{\rm GR}_{\overline{\nu}^{}_{e}e} \simeq 4.86 \times 10^{-31}~{\rm cm^2}$ when $\sqrt{s} = M^{}_{W} \simeq 80.4~{\rm GeV}$ \cite{Tanabashi:2018oca}. Since electrons are basically static inside the Earth, a resonance takes place with the neutrino energy $E^{}_{\overline{\nu}_{e}} \simeq 6.3~{\rm PeV}$. In Fig.~\ref{fig:xsec}, we compare the cross sections of different scattering processes.
The cross sections of DIS are calculated using CTEQ6 parton distribution functions \cite{edsjo2007calculation}. 
We notice that in a wide energy range from $4~{\rm PeV}$ to $10~{\rm PeV}$, the cross section of $\overline{\nu}^{}_{e}{-}e$ scattering exceeds that of DIS. Here we ignore the subdominant scattering processes \cite{Seckel:1997kk,Alikhanov:2014uja,Alikhanov:2015kla,Gauld:2019pgt,Zhou:2019vxt,Beacom:2019pzs} and the motions of atomic electrons \cite{Loewy:2014zva} for simplicity. 

Assuming the Earth matter is composed of the standard rock with the density $\rho = 2.65~{\rm g \cdot cm^{-3}}$ and the electron fraction $50\%$, we can figure out the MFPs of neutrinos at $E^{}_{\nu} = 6.3~{\rm PeV}$ propagating inside Earth using the formula $\lambda = ( \sigma\cdot n^{}_{\rm tar})^{-1}$:
\begin{align}
\lambda^{\rm GR}_{\overline{\nu}^{}_{e}e} \simeq 26~{\rm km},~
\lambda^{\rm CC}_{\overline{\nu}^{}_{e}N} \simeq 4436~{\rm km}\;,
\label{eq:MFP}
\end{align}
for the Glashow resonance channel and CCDIS with nucleons, respectively.
Therefore, in the energy band of $4 {-} 10~{\rm PeV}$, the MFP of $\overline{\nu}^{}_{e}$-$e$ scattering dominates and lies in the range of $26 {-} 4436 ~{\rm km}$. 
The MFPs can be simply rescaled if a different matter configuration is considered.
The decay of $W$ produces either hadronic showers ($\sim 67\%$) or leptons directly ($\sim 11\%$ for each flavor). The energetic final state of the electron easily ends up into an electromagnetic shower in the medium. Meanwhile, the produced muon will survive as a long track. 
The tau has very short lifetime and decays very fast with a decay length $\lambda_{\tau}\simeq 50\,\mathrm{m}\cdot(E_\tau/{\rm PeV})$.

The production of muons and taus by astrophysical neutrinos is of our main concern. To obtain the fluxes of muons and taus near the detector location, we need to first evolve the neutrino flux in the medium given some initial injection spectra.
The evolution equation of neutrinos propagating inside the Earth can be expressed as
\begin{equation}
\frac{\partial\rho}{\partial r} = -i[H, \rho]-\{\Gamma,\rho\}+\int_E^\infty f(\rho,\overline{\rho},r,E',E)dE',
\label{eq:evol}
\end{equation}
where $\rho$ is the density matrix of the neutrino state $\left|\Psi^{}_{\nu} \right> \left< \Psi^{}_{\nu} \right|$, which can be written as a $3{\times}3$ matrix in the flavor basis.
On the right hand side, the first term describes neutrino oscillations including the matter effect. The oscillation effect actually becomes negligible at high energies, e.g., the oscillation length $L^{31}_{\rm osc} \simeq 10^7~{\rm km}\cdot \left[E^{}_{\nu}/\left(10~{\rm TeV}\right)\right]$ corresponding to the larger mass squared difference far exceeds the Earth diameter. The second term signifies the attenuation of neutrinos in the Earth. The damping matrix $\Gamma$ contains the DIS with nucleons as well as the scattering with electrons including the Glashow resonance channel.
The last term stands for the neutrino yield as the interaction products of neutrinos at higher energies, including tau regeneration and $W$ decay. All these effects are taken into account for the neutrino propagation with the help of {nuSQuIDs} \cite{delgado2016nusquids}.
Here, we use the Preliminary Reference Earth Model (PREM) as the density profile of the Earth \cite{Dziewonski:1981xy} and an isoscalar Earth target is assumed.

\section{Through-going Events at IceCube}
The IceCube observatory is located at the South Pole and comprised of a $1~\rm{km^3}$ in-ice array starting from $1.45~{\rm km}$ underground \cite{Aartsen:2016nxy}. The detecting modules are surrounded by ice with the Antarctic rock beneath the detector volume. With a cubic kilometer of ice target, IceCube is sensitive to high-energy neutrinos through the detection of Cherenkov light emitted from the secondary charged particles.
%\begin{align}
%\frac{{\rm d}^2\Phi^{}_{{6\nu}}}{{\rm d} E_{\nu}  \mathrm{d}\Omega} =\, A \left(\frac{E^{}_{\nu}}{100~{\rm TeV}} \right)^{-\gamma} \!\! \cdot 10^{-18}  ~{\rm GeV^{-1}cm^{-2}s^{-1}sr^{-1}} \;
%\label{eq:ICFit}
%\end{align}
%for all neutrino flavors, where the normalization factor $A=6.45^{+1.46}_{-0.46}$ and the spectrum index $\gamma=2.89^{+0.19}_{-0.2}$ have been obtained by fitting 7.5 years of IceCube HESE data \cite{Schneider:2019ayi}, and $A=4.32^{+0.75}_{-0.72}$ and $\gamma = 2.28^{+0.09}_{-0.08}$ with ten years of TGM data \cite{Stettner:2019tok}. 
The differential spectrum of muons produced by the Glashow resonance as a function of the zenith angle $\theta$ and the muon energy $E^{}_{\mu}$ when entering the IceCube detector can be written as
\begin{align}
%\frac{\mathrm{d}^2\Phi^\mathrm{}_{\overline{\nu}^{}_{e}}}{\mathrm{d} E_{\overline{\nu}^{}_{e}} \mathrm{d}\Omega}
&\Phi^{\mathrm{GR}}_\mu(\theta, E_\mu) = \int\int \mathrm{d}x\,\mathrm{d}E_{\overline{\nu}_{e}}\Phi^{}_{\overline{\nu}_e}(E_{\overline{\nu}^{}_{e}}, \theta, x) \nonumber \\
& \times \int \mathrm{d}E^{\prime}_\mu \, \sigma^{}_{\overline{\nu}^{}_{e}e}\, \frac{1}{\Gamma_{W}}\frac{\mathrm{d}\Gamma_{W \rightarrow\mu}}{\mathrm{d}E_\mu'} \,  n^{}_e(\theta,x) \cdot \frac{\mathrm{d}\mathcal{P}}{\mathrm{d}E^{}_\mu}(\theta,x) \;,
\label{eq:mu_flux}
\end{align}
where $x$ stands for the distance from the interaction vertex to the detector, $n^{}_e$ is the electron number density at $x$, 
%${\rm{d}\sigma^{ \mu}_{\overline{\nu}^{}_{e}e}}/{\rm{d}E_\mu'}$ is the 
$\mathrm{d}\Gamma_{W\rightarrow\mu}/\mathrm{d}E_\mu'$ is the decay rate of $W$ to a muon with the energy $E^{\prime}_{\mu}$,
and the probability function ${\mathrm{d}\mathcal{P}}/{\mathrm{d}E^{}_\mu}$ links 
$E_\mu'$ at the production point to $E_\mu$ at the detector location after considering the energy loss of the muon.
The muon flux from DIS with nucleons will also contribute to through-going track events, as a background, and it is similarly given by
\begin{align}
&\Phi^{\mathrm{CC}}_{\mu,1}(\theta, E_\mu)= 
 \sum^{}_{N=p,n}\int\int \mathrm{d}x\,\mathrm{d}E^{}_{\nu^{}_{\mu}}\Phi_{{\nu}_\mu}(E^{}_{\nu^{}_{\mu}}, \theta, x) \nonumber\\
& \times  \int \mathrm{d}E^{\prime}_\mu \, \frac{\mathrm{d}\sigma^{ \mu}_{{\nu}^{}_{\mu}N}}{\mathrm{d}E_\mu'} \,  n^{}_N(\theta,x) \cdot \frac{\mathrm{d}\mathcal{P}}{\mathrm{d}E^{}_\mu}(\theta,x) \;, 
\label{eq:CC_flux_mu}
\end{align}
\begin{align}
&\Phi^{\mathrm{CC}}_{\mu,2}(\theta, E_\mu) = 
\sum^{}_{N=p,n}\int\int \mathrm{d}x\,\mathrm{d}E^{}_{\nu^{}_{\tau}}\Phi^{}_{{\nu}_\tau}(E^{}_{\nu^{}_{\tau}}, \theta, x) \nonumber\\
& \int  \int \mathrm{d}E^{\prime}_\mu \mathrm{d}E^{}_{\tau} \frac{\mathrm{d}\sigma^{ \tau}_{{\nu}^{}_{\tau}N}}{\mathrm{d}E^{}_{\tau}} 
 \frac{\mathrm{d}\Gamma_{\tau \rightarrow \mu}}{\Gamma_{\tau} \mathrm{d}E_\mu'} \, n^{}_N(\theta,x) \cdot \,
 \frac{\mathrm{d}\mathcal{P}}{\mathrm{d}E^{}_\mu}(\theta,x) \;,
\label{eq:CC_flux_tau}
\end{align}
where both the direct production from the CC interction of muon neutrinos and the decay from secondary tau are taken into account. Here, $\mathrm{d}\sigma^{}_{\nu N}/\mathrm{d}E^{}_l$ is the differential CC cross section at a charged-lepton energy $E^{}_l$ with $N$ being either proton or neutron, $n^{}_N$ the number density of the target nucleon, and $\mathrm{d}\Gamma^{}_{\tau \rightarrow \mu}/\mathrm{d}E^{\prime}_\mu$ the decay rate of tau to muon with the energy $E^{\prime}_\mu$. 
The energy loss of taus is negligible until $\sim 10^9$ GeV in ice when the interaction length becomes comparable to the decay length \cite{koehne2013proposal}. Therefore, it is safe to assume an instant decay of taus at the spot of creation in this case.     
\begin{figure}[t!]
	\centering
	\hspace{-0.5cm}
	\includegraphics[width=0.84\columnwidth]{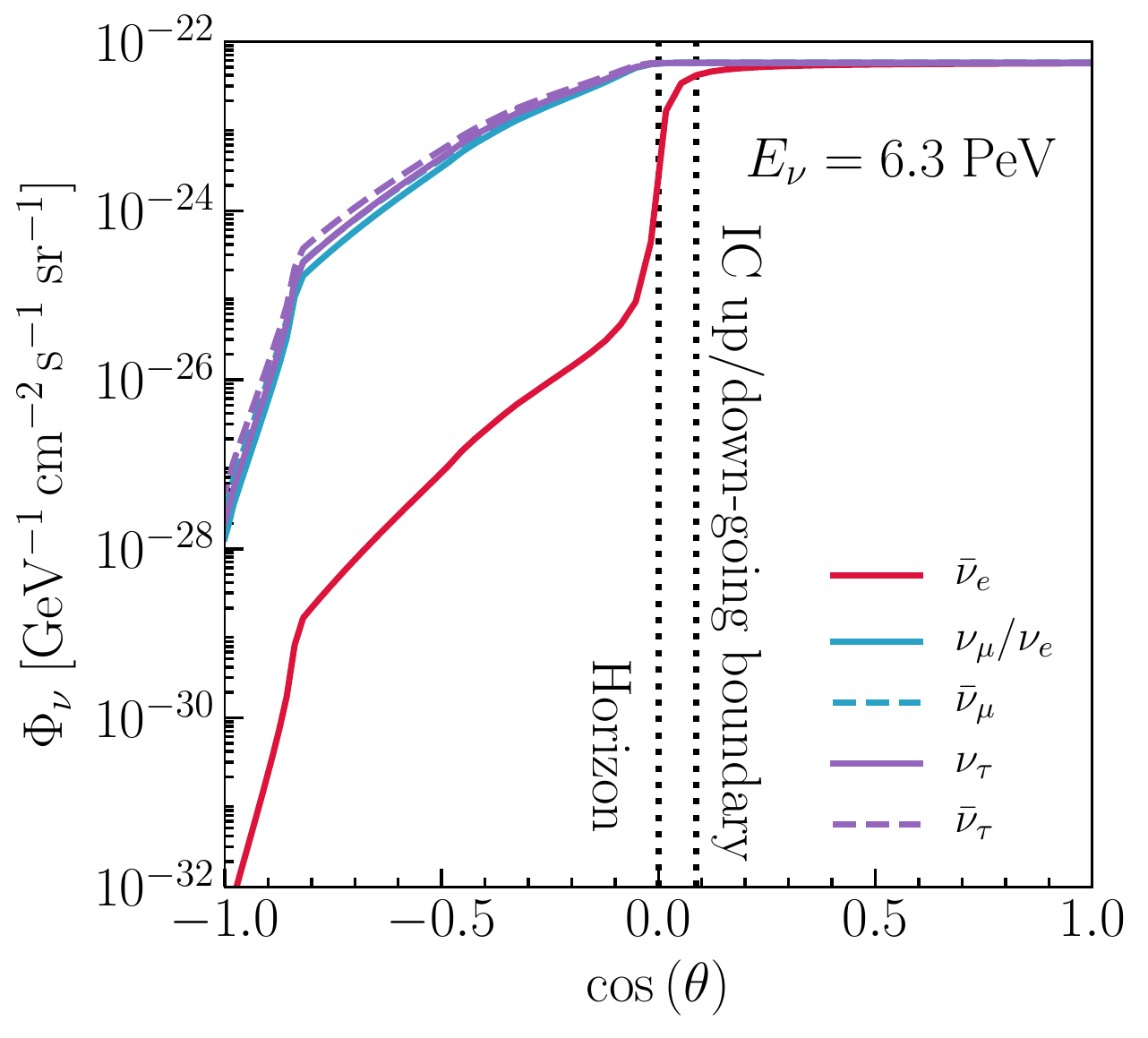}
	\caption{Zenith distribution of neutrino fluxes of all flavors at the Glashow resonant energy when the distance is 1 km away from the center of the IceCube detector. The horizon and the zenith cuts of the northern/southern sky for IceCube are shown as black dotted lines. \label{fig:nu_flux}}
\end{figure}

The IceCube experiment measures the diffuse astrophysical neutrino spectrum and usually an unbroken power-law spectrum is assumed in the fit. The most recent analysis of TGM from the Northern Hemisphere with 10 years data yields a spectrum index $\gamma = 2.28$ \cite{Stettner:2019tok}. Here, we start from the latest TGM spectrum analysis and set the flavor composition of injected neutrinos as $1:1:1$ with equal amounts for neutrinos and antineutrinos for the purpose of demonstration. The neutrino flux injected into the Earth is assumed to be isotropic. Fig.~\ref{fig:nu_flux} shows the differential neutrino fluxes of all flavors at $E^{}_{\nu} = 6.3~{\rm PeV}$ when the distance of neutrinos to the IceCube detector is $1~{\rm km}$. Significant attenuation effects  can be observed, especially for the case of $\overline{\nu}_e$ due to the resonant scattering. The flux of $\overline{\nu}_e$ starts attenuating even from the angle above the horizon with $\cos{\theta} > 0$, while the attenuation effect of other flavors gradually increases after $\cos{\theta} < 0$. This can be easily understood from their respective MFPs. The MFPs in Eq.~(\ref{eq:MFP}) correspond to the zenith angles $\cos{\theta}_{\rm{GR}} \sim 0.04$ for the Glashow resonance and $\cos{\theta}_{\rm{CC}} \sim -0.35$ for CCDIS. Slightly below the horizon, the $\overline{\nu}^{}_{e}$ flux at the resonant energy starts to be significantly absorbed by the Earth matter, and only a very small fraction can reach the IceCube detector. Therefore we expect that the through-going events from the Glashow resonant scattering mainly come from around the horizontal direction.

After being produced, the charged lepton loses its energy during the propagation to the detector through various processes: ionization, bremsstrahlung, photo-nuclear interactions and electron pair production. The propagation code {PROPOSAL} \cite{koehne2013proposal} is used to simulate energy losses of muons. The interaction length of muons at PeV energies in the Earth matter is on the order of kilometer.
Only a relatively small volume surrounding the detector is relevant for the muon production at our concerned energies.  
We set a maximum distance of $25~{\rm km}$ from the detector to integrate and yield the secondary charged lepton flux. It is unlikely that muons produced beyond this range are able to be detected due to the attenuation.   

%\begin{align}
%D(\theta) = &~ \left|\cos{\theta} \right|    \sqrt{R^{2}_{\oplus} - h(h-2R^{}_{\oplus}) \tan^2{\theta}}   \nonumber \\
%& + \cos{\theta} \cdot \left( h-R^{}_{\oplus}\right)  \;, \nonumber\\
%\cos{\theta} = &~ \frac{D^2+h^2-2h R^{}_{\rm \oplus}}{2D h - 2 D R^{}_{\rm \oplus}} \;.
% \label{eq:MFP}
%\end{align}
%However, as we noted before under Eq.~(\ref{eq:MFP}), the residue of the resonant enhancement spans a very wide energy range of  $\left(4 \cdots 10 \right)~{\rm PeV}$. Neutrinos in this energy range will have an attenuation behavior in between those of $\overline{\nu}^{}_{e}$ and other flavors in Fig.~\ref{fig:nu_flux}, and the final results should be their average.

We may first make a rough estimation of the through-going events at IceCube.
Assuming a constant density profile for Earth, for each neutrino energy, we compute its typical injection zenith angle $\theta^{}_{\rm MFP}$ corresponding to the MFP. The produced angular-integrated muon flux can be obtained by
$
{\mathrm{d}N^{}_{\mu}}/{\mathrm{d}E^{}_{\mu}} = \Phi_\nu \cdot \Delta\Omega \cdot \mathcal{P}^{}_{\nu \rightarrow \mu} \;,
$
where $\mathcal{P}^{}_{\nu \rightarrow \mu}$ represents the probability that a neutrino can produce a detectable $\mu$ at the detector, and $\Delta \Omega$ stands for the contributing solid angle. The probability can be estimated with $\mathcal{P}^{}_{\nu \rightarrow \mu} \sim \lambda^{}_{\mu}/\lambda^{}_{\nu}$, where $\lambda^{}_{\mu}$ is the survival length scale of muons. This distance may be parameterized with $\lambda^{}_{\mu} = 1/(b \rho) \cdot \ln\left(1+E^{}_{\mu} \cdot b/a\right)$ with $b \simeq 4.64 \times 10^{-6}~{\rm g^{-1}\cdot cm^{-2}}$, $a \simeq 2.22 \times 10^{-3}~{\rm GeV\cdot g^{-1}\cdot cm^{-2}}$ and $\rho = 2.65~{\rm g\cdot cm^{-3}}$ for the standard rock \cite{Lipari:1991ut,koehne2013proposal}. 
For $E^{}_{\mu} > 10~{\rm TeV}$, the radiation losses will dominate over the ionization losses, and the survival distance will follow a simple relation $\lambda^{}_{\mu} \sim 4~{\rm km} + 2~{\rm km} \cdot \log^{}_{10} \left[ E/(100~{\rm TeV)} \right]$ for standard rock and $\lambda^{}_{\mu} \sim 15~{\rm km} + 7~{\rm km} \cdot \log^{}_{10} \left[ E/(100~{\rm TeV)} \right]$ for ice.
The solid angle can be obtained with $\Delta \Omega \sim 2\pi   \cdot\left|\cos{\theta}^{}_{\rm MFP} - \cos{\theta^{}_{\rm GR}}\right| $.
We then estimate the event number by using 
$
N^{}_{\mu} = \int \mathrm{d}E^{}_{\mu}\,{\mathrm{d}N^{}_{\mu}}/{\mathrm{d}E^{}_{\mu}} \cdot A^{}_{\rm geo} \cdot T\;,
$
where for this estimation the geometrical area $A^{}_{\rm geo}$ is fixed to $1~{\rm km^2}$, and $T \sim 1~{\rm yr}$ is the exposure. The number of Glashow resonance events is obtained as $N^{}_{\mu} \simeq 0.15$ with an input of the TGM spectrum and $N^{}_{\mu} \simeq 0.025$ with the best-fit HESE spectrum \cite{Schneider:2019ayi}. 

There are two types of backgrounds for the Glashow resonance signature: (i) the muon generated through CCDIS with nucleons; (ii) the atmospheric muon background produced by high-energy cosmic rays.
The atmospheric background is relevant when the muon is coming from the southern sky, while the northern sky of IceCube is basically shielded by the Earth. We take account of both the conventional atmospheric muon flux from the full shower CORSIKA simulation \cite{fedynitch2012influence} and the prompt one from theoretical estimations \cite{enberg2008prompt} assuming an isotropic distribution.

\begin{figure}[t!]
	\centering
	\includegraphics[width=0.9\columnwidth]{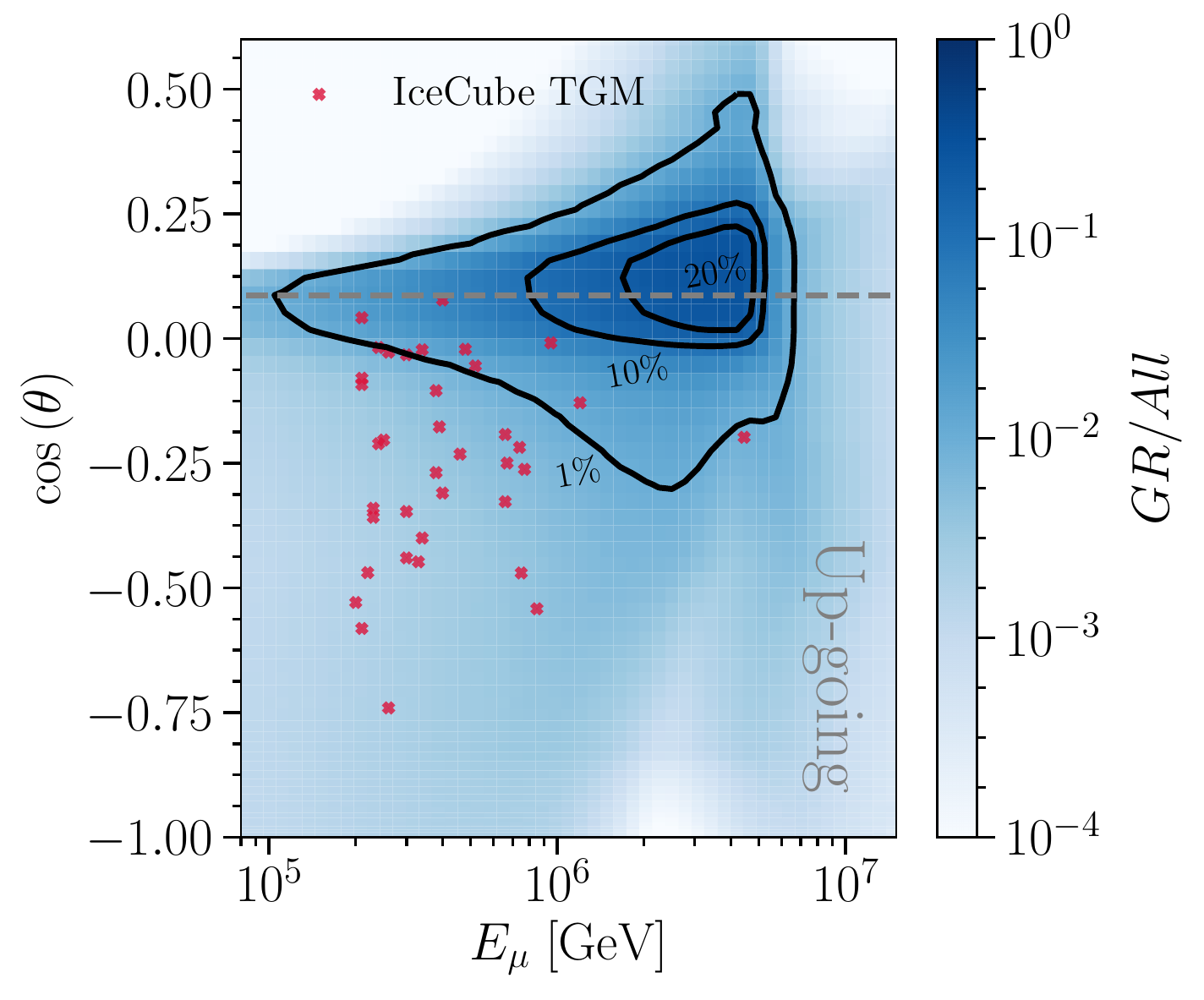}
	\caption{The fraction of  through-going events induced by the Glashow  resonance as a function of the zenith angle and muon entering energy. The total events include the Glashow resonance channel, the charged current deep inelastic scattering channel and the atmospheric muon background. The available IceCube  through-going muon events of Northern Hemisphere with proxy energy higher than 200 TeV are shown as red crosses \cite{Aartsen:2016xlq,haack2018measurement,Stettner:2019tok}.  The horizontal dashed line stands for the applied zenith angle cut of $85^{\circ}_{}$ for these events. \label{fig:IC_GR_ratio}}
\end{figure}

\begin{figure*}[t!]
	\centering
	\hspace{-1cm}
	\subfigure{\includegraphics[width=0.82\columnwidth]{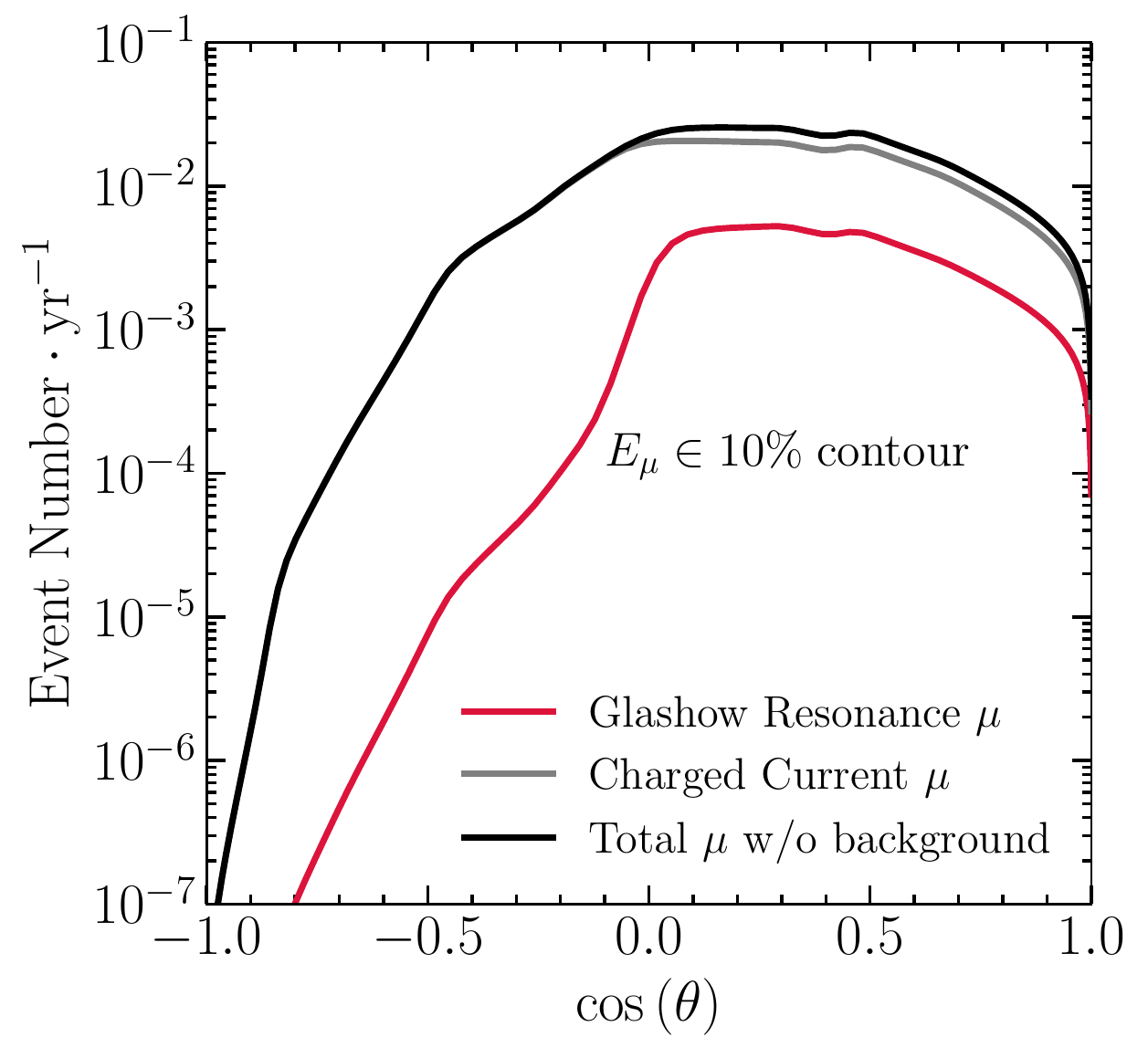}} 	\hspace{1cm}
	\subfigure{\includegraphics[width=0.81\columnwidth]{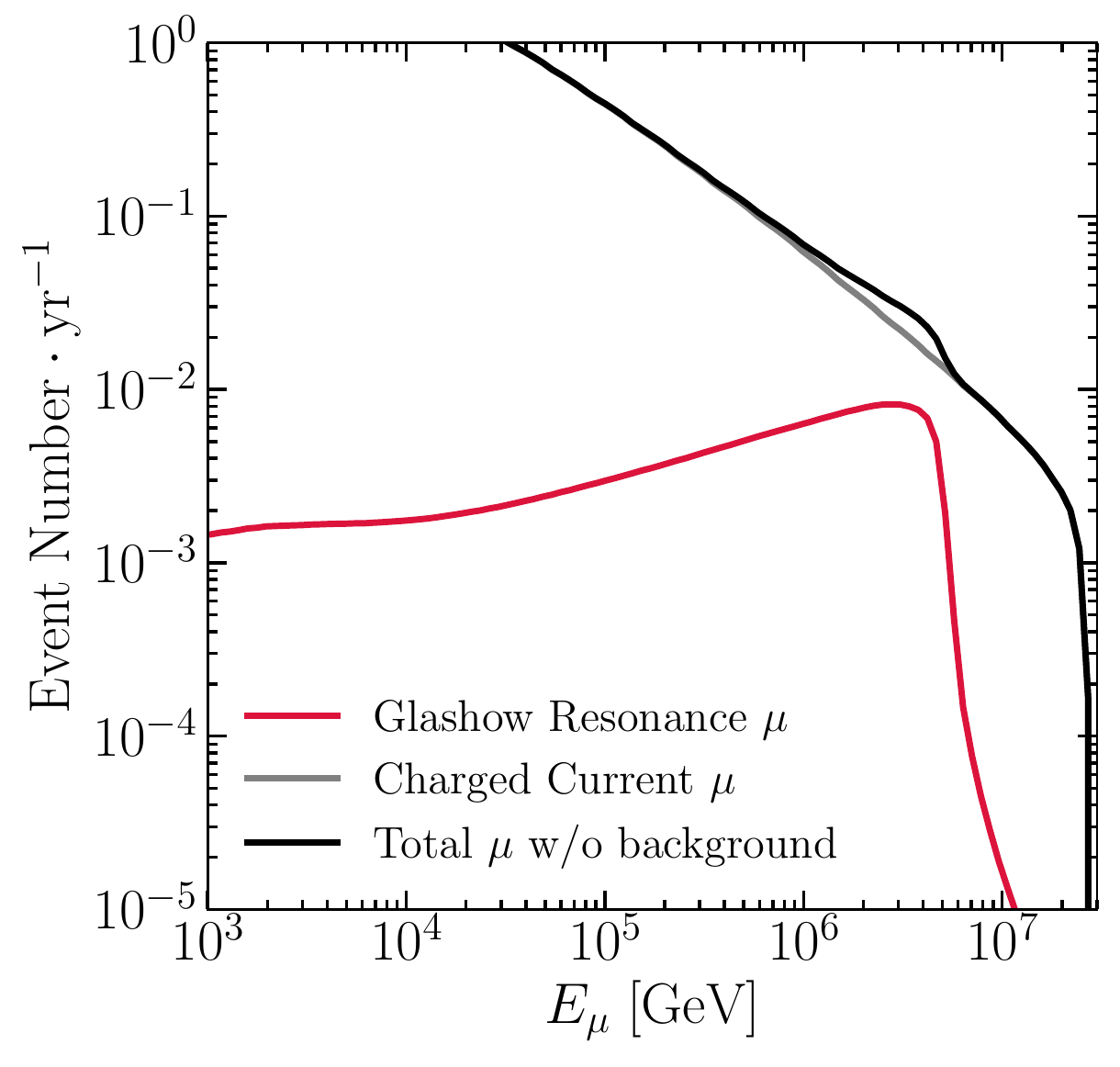}}
	\caption{The event number per year for through-going muons at IceCube as functions of the zenith angle (left panel) and the muon energy (right panel). The atmospheric  background is not included here.
	In the left panel, the Glashow resonance excess arises when the incoming zenith angle is near the horizon. The event excess in the southern sky ($\cos{\theta} > 0$) is easily contaminated by the atmospheric background. In the right panel, an excess at energies around PeV can be noticed.\label{fig:IC_event}}
\end{figure*}

In Fig.~\ref{fig:IC_GR_ratio}, as a function of the muon energy and the incoming zenith angle, we show the fraction of through-going events induced by the Glashow resonance to the total events, including the Glashow resonance channel, the CCDIS channel as well as the atmospheric muon background. A clear excess of events around the horizon in the PeV energy region can be noticed. The contours inside which the fraction of Glashow resonance events over all events is larger than the indicated value are shown. The total number of the Glashow resonance events is 0.3 and 0.028 within the $10\%$-contour for one year of data taking. Fig.~\ref{fig:IC_event} shows the distribution of the Glashow resonance events with respect to the muon energy and incoming zenith angle, respectively. In the left panel, we have marginalized the muon energy in the range of the $10\%$-contour, while in the right panel, the muon zenith angle has been marginalized. 
With ten years of event accumulation, only 0.28 Glashow resonance events are expected within the $10\%$-contour excessive parameter region. The future IceCube-Gen2 upgrade will increase the muon event rate by a factor of about 4 around the horizon with an increase of volume by a factor of nearly ten \cite{Ackermann:2017pja}. For ten years data taking with IceCube-Gen2, one expects the event number to be 1.12 within the contour.

\section{Earth-skimming Neutrinos}
By lifting a telescope to high altitude, which could be on the top of a mountain, attached to a balloon in flight, or floating in outer space with a satellite, a large area close to the horizon of the Earth can be overwatched. 
As mentioned before, due to the attenuation effect of Earth matter, neutrinos with large cross sections can be observed only when they are coming from an angle slightly below the Earth horizon. The decay of secondary tau from Earth-skimming neutrinos induces a shower signal, which can be detected by those telescopes with various techniques \cite{Alvarez-Muniz:2018bhp,Cao:2004sd,Gorham:2016zah,Ahnen:2018ocv,Allison:2018cxu,Aab:2019auo,Aguilar:2019jay,Otte:2018uxj,Otte:2019aaf,Alvarez-Muniz:2018bhp,Neronov:2016zou,Krizmanic:2019hiq,Neronov:2019htv,Sasaki:2014mwa}.
Among them, telescopes with the Cherenkov and fluorescence techniques are ready to reach an energy threshold as low as PeV, e.g., MAGIC \cite{Ahnen:2018ocv}, NTA \cite{Sasaki:2014mwa}, POEMMA \cite{Krizmanic:2019hiq}, Trinity \cite{Otte:2018uxj,Otte:2019aaf}, and CHANT \cite{Neronov:2016zou}. They can be used to probe $\overline{\nu}^{}_{e}$ via the Glashow resonance production of tau.

We take CHANT as an example to illustrate the sensitivity of this type to the $\overline{\nu}^{}_{e}$ detection via the Glashow resonance channel. CHANT is a proposed space-borne imaging atmospheric Cherenkov telescope (IACT) which can achieve  the most competitive sensitivity among similar proposals in the neutrino energy range of PeV-EeV \cite{Neronov:2016zou}. 
The high altitude of the outer-space telescope ($\sim 300~{\rm km}$) enables the detection of high-energy neutrinos with a very wide field of view. Almost the entire near-horizontal part of the Earth can be treated as the target medium for neutrino interactions.

The decay length of tau at PeV energies is very short, as mentioned previously, so we expect the conversion probability of neutrinos into taus $\mathcal{P}^{}_{\nu \rightarrow \tau}$ to be proportional to $\lambda_{\tau}/\lambda_{\nu}$. It seems that with a larger cross section (smaller $\lambda_{\nu}$), the conversion efficiency will be larger, such that the final event rate is also larger. Since the cross section of the Glashow resonance is much larger than the CCDIS one around the resonant energy, one may expect a significant enhancement of the $\overline{\nu}^{}_{e}$ events. However, very similar to the TGM at IceCube, this turns out not to be the case in the Earth-skimming scenario, and the reason is as follows.
The final event rate should be proportional to a combination of several terms $N^{}_{\tau} \propto \mathcal{P}^{}_{\nu \rightarrow \tau} \cdot \Delta \Omega \cdot A^{}_{\rm geo}$ with $ A^{}_{\rm geo}$ signifying the detector acceptance of Cherenkov emission after tau decays, and $\Delta \Omega$ being the contributing solid angle from the satellite. $\Delta \Omega$ is determined by the critical zenith angle, at which the chord length of neutrino trajectory inside Earth is equal to the neutrino MFP. 
At the exact Glashow resonance energy, the MFP of $\overline{\nu}^{}_{e}$ in Eq.~(\ref{eq:MFP}) corresponds to an extremely narrow angle below the horizon, i.e., $0.0004^{\circ}$.
It is easy to verify that $\Delta \Omega \propto \lambda^{2}_{\nu}$. Since $\lambda^{}_{\tau}$ is fixed at the resonant energy, the final event rate scales as $N^{}_{\tau} \propto \lambda^{}_{\nu}$.
On the other hand, for CCDIS, as the neutrino energy increases, the available target volume of the Earth reduces in the same manner.
But this will be compensated by a boost of the interaction length of tau $\lambda^{}_{\tau}$.
The geometric area  $A^{}_{\rm geo}$ will also increase due to a stronger Cherenkov signal for a higher neutrino energy. In the following we justify the qualitative discussion here with a numerical calculation.

\begin{figure}[t]
	\centering
	\includegraphics[width=0.9\columnwidth]{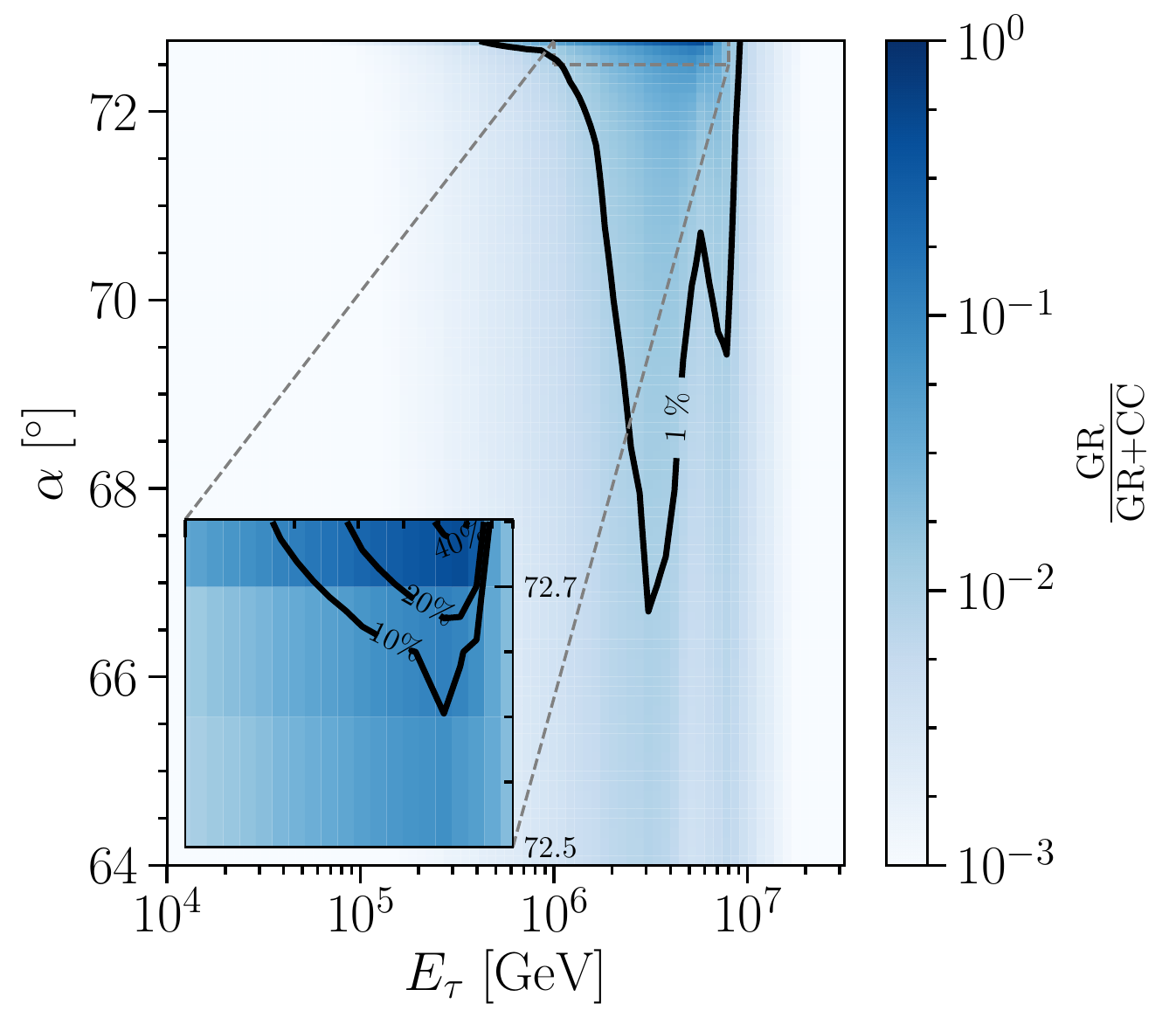}
	\caption{The ratio of shower events induced by $\overline{\nu}^{}_{e}$ via the Glashow resonance to the event sum of $\overline{\nu}^{}_{e}$ and ${\nu}^{}_{\tau}$+$\overline{\nu}^{}_{\tau}$ as a function of the elevation angle and emerging tau energy, in a CHANT-like experiment. In the  bottom left corner, the region of the energy range 1-8 PeV near the Earth horizon is zoomed in.  \label{fig:chant_GR_ratio}}
\end{figure}

The detectable $\overline{\nu}^{}_{e}$-induced tau flux through the Glashow resonance channel is given by
\begin{align}
\Phi^{\mathrm{GR}}_\tau(\theta, E^{}_{\tau}) = 
\int \mathrm{d}E_{\overline{\nu}_{e}}\Phi^{\mathrm{sf}}_{\overline{\nu}_e}
\frac{\mathrm{d}\sigma^{ \tau}_{\overline{\nu}^{}_{e}e}}{\mathrm{d}E^{}_{\tau}} n^{}_{e} \cdot \lambda_{\tau},
\label{eq:taufluxGR}
\end{align}
where $\Phi^\mathrm{sf}_{\overline{\nu}^{}_{e}}$ stands for the $\overline{\nu}^{}_{e}$ flux near the Earth surface. An approximation has been made in deriving Eq.~(\ref{eq:taufluxGR}) that the energy loss of tau can be ignored and the decay length of tau is very short in this energy range as discussed previously.
Those taus are basically produced within $\lambda_{\tau}$ on average  beneath the Earth surface, and then decay in the atmosphere to produce EAS.   
On the other hand, taus produced in CCDIS of $\nu^{}_{\tau}$ is a background for the Glashow resonance event search, and its flux reads
\begin{align}
\Phi^{\mathrm{CC}}_\tau(\theta, E^{}_{\tau}) = 
\sum^{}_{N=p,n} \int \mathrm{d}E_{\nu^{}_{\tau}}\Phi^{\mathrm{sf}}_{{\nu}_\tau}
\frac{\mathrm{d}\sigma^{ \tau}_{\nu^{}_{\tau} N}}{\mathrm{d}E^{}_{\tau}} n^{}_{N} \cdot \lambda_{\tau},
\label{eq:taufluxCC}
\end{align}
where $\sigma^{ \tau}_{\nu^{}_{\tau} N}$ is the cross section for  $\tau$ production, and other notations follow the definitions in previous equations.
Setting the telescope to an altitude of $300~{\rm km}$, in Fig.~\ref{fig:chant_GR_ratio} we show the ratio of tau events from the Glashow resonance to the sum of all contributing channels. 
The elevation angle $\alpha$ of incoming neutrinos with respect to the vertical line has been defined, with $\alpha_{\rm max}\approx 72.75^\circ$ when the neutrino trajectory is tangent to the Earth surface.
The excess due to the Glashow resonance is mainly in the region close to the horizon within a very narrow angle $\sim 0.1^{\circ}$. 
It is technically viable to resolve the events in this narrow region, given that IACT has a very good angular resolution $\sim 0.1^\circ$.
However, if we have the IceCube TGM best-fit spectrum as the input,
with one year of exposure the event number of only 0.024 is expected within the $10\%$-contour. It is thus very difficult to capture the Glashow resonance signature by observing the Earth-skimming neutrinos. 

Another detection mechanism complementary to the Earth-skimming technique is to observe the shower directly induced by a deeply penetrating neutrino in the atmosphere \cite{Berezinsky:1975zz,Aab:2015kma,Alvarez-Muniz:2018bhp}. This method is sensitive to neutrinos of all flavors. However, to reject the cosmic ray background, it requires some additional techniques to discriminate between the `old' showers initiated by cosmic rays and `young' showers by neutrinos. The efficiency will also drop due to the selection criteria.

\section{Mountain as a Target}
A high mountain can also be treated  as a good target for neutrino interactions \cite{Fargion:1999se,Fargion:2000iz,Sasaki:2014mwa,Aita:2011mx,Vannucci:2001am,Hou:2002bh,Asaoka:2012em,Gora:2014lya,Gora:2016mmy}. Neutrino-induced tau signals can be detected by placing a telescope on one side of a valley opposite to a mountain. The mountain in this case acts both as  a filter of cosmic rays and a target volume for neutrino interactions.
One advantage of these kind of experiments is that the detection volume will not shrink like the previous cases as the neutrino cross section increases.

There are several on-going and planned experiments advancing towards this direction, e.g., GRAND \cite{Alvarez-Muniz:2018bhp}, Ashra \cite{Aita:2011mx,Asaoka:2012em} and NTA \cite{Sasaki:2014mwa}.
\begin{figure}[t]
	\centering
	\hspace{-0.8cm}
	\includegraphics[width=0.82\columnwidth]{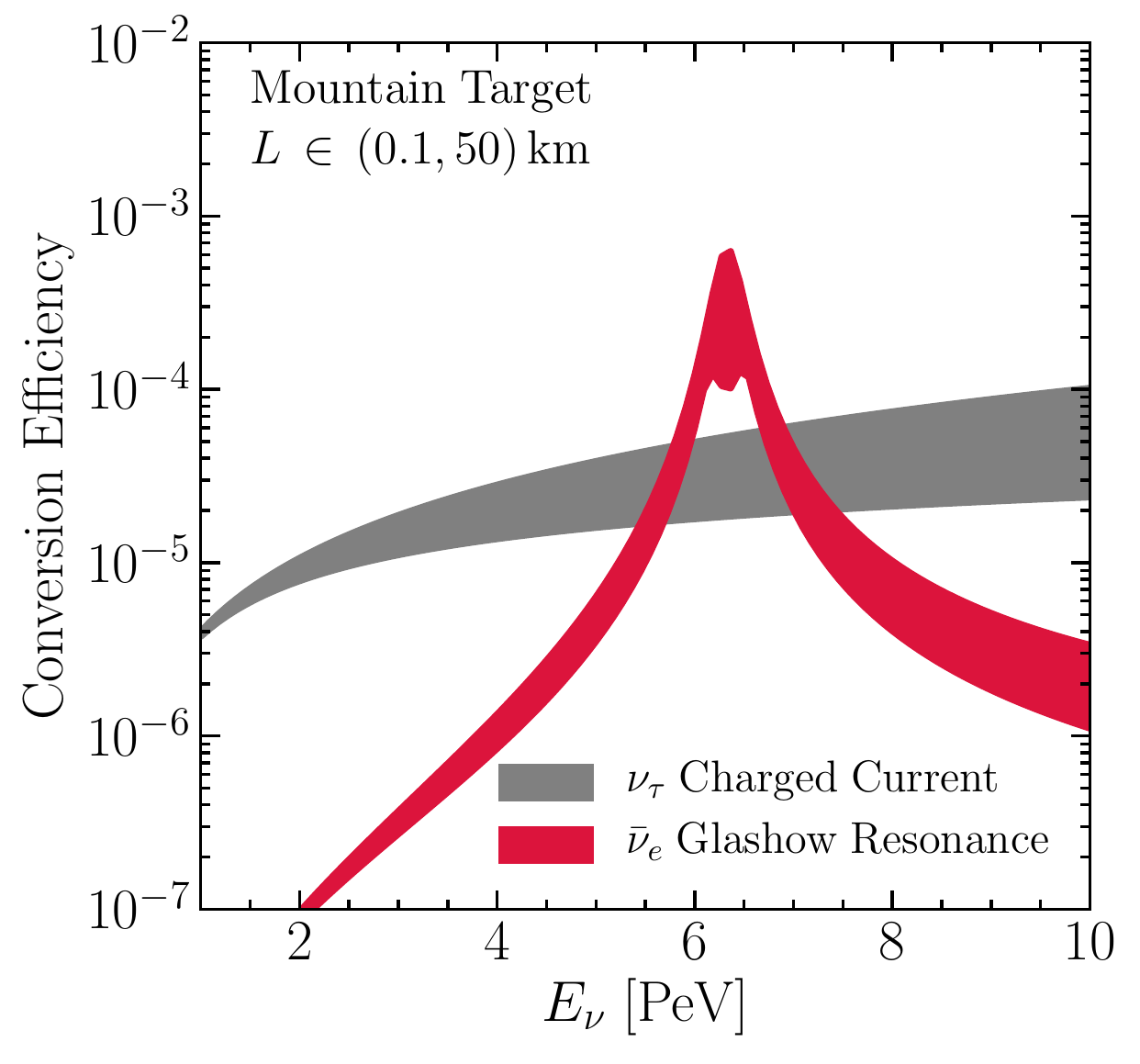}
	\caption{The conversion efficiency from a mountain-penetrating neutrino ($\overline{\nu}^{}_{e}$ or $\nu^{}_{\tau}$) to an emerging tau at the mountain surface. The tau production via the Glashow resonance channel is shown as the red band, while that through CCDIS of $\nu^{}_{\tau}$ as the gray band. The band is obtained by varying the mountain thickness from 0.1 km to 50 km. The mountain is assumed to be composed of standard rock. \label{fig:MTCE}}
\end{figure}
The GRAND experiment will use the radio array to monitor the mountain. However, the minimal energy that GRAND is sensitive to is limited by the triggering threshold of the antennas. It might be challenging for the detection of neutrinos with energies down to the Glashow resonance at GRAND. The Ashra experiment using the atmospheric Cherenkov technique, has set a limit on the neutrino flux from point sources in the PeV-EeV energy range, by overlooking the Mauna Kea on Hawaii Island \cite{Aita:2011mx}. NTA is a Cherenkov-fluorescence combined telescope on Hawaii Island evolved from Ashra. We shall focus on the latter case.

To illustrate the idea and make the result as transparent as possible, the mountain will firstly be simplified to a block with the thickness of $L$ following Ref.~\cite{Hou:2002bh}. Neutrinos are injected into this nominal mountain from one side, interacting with the standard rock and generating observable taus which emerge from the other side of the mountain wall. The conversion efficiency from an injected $\overline{\nu}^{}_{e}$ into a tau can be estimated with 
\small
\begin{align}
\mathcal{P}^{\rm GR}_{\nu \rightarrow \tau} = &~  \frac{\lambda^{}_{\tau}}{\lambda^{}_{\overline{\nu}^{}_{e}}-\lambda^{}_{\tau}} \left( \mathrm{e}^{-L/\lambda^{}_{\overline{\nu}^{}_{e}}} - \mathrm{e}^{-L/\lambda^{}_{\tau}} \right)  \frac{\lambda^{}_{\overline{\nu}^{}_{e}}}{\lambda^{\rm GR}_{\overline{\nu}^{}_{e}e}} \times 11 \% \;,
\label{eq:efficiencyGR}
\end{align}
\normalsize
where $\lambda^{}_{\overline{\nu}^{}_{e}} \equiv \left( 1/\lambda^{\rm GR}_{\overline{\nu}^{}_{e}e} + 1/\lambda^{\rm CC}_{\overline{\nu}^{}_{e}N} + 1/\lambda^{\rm NC}_{\overline{\nu}^{}_{e}N} \right)^{-1}$ is the total MFP of $\overline{\nu}^{}_{e}$, the fraction ${\lambda^{}_{\overline{\nu}^{}_{e}}}/{\lambda^{\rm GR}_{\overline{\nu}^{}_{e}e}}$ severely depends on the neutrino energy around the Glashow resonance point, and ${\rm Br}(W^{} \rightarrow \tau + \nu^{}_{\tau}) \simeq 11\%$. On the other hand, the conversion efficiency of $\nu^{}_{\tau}$ to a detectable tau is given by \cite{Hou:2002bh}
\begin{align}
\mathcal{P}^{\rm CC}_{\nu \rightarrow \tau} = &~  \frac{\lambda^{}_{\tau}}{\lambda^{}_{\nu^{}_{\tau}}-\lambda^{}_{\tau}} \left( \mathrm{e}^{-L/\lambda^{}_{\nu^{}_{\tau}}} - \mathrm{e}^{-L/\lambda^{}_{\tau}} \right) \frac{\lambda^{}_{\nu^{}_{\tau}}}{\lambda^{\rm CC}_{\nu^{}_{\tau}N}},
\label{eq:efficiencyCC}
\end{align}
with ${\lambda^{}_{\nu^{}_{\tau}}}/{\lambda^{\rm CC}_{\nu^{}_{\tau}N}} \simeq 1$ because CC interaction dominates.

\begin{figure*}[t!]
	\centering
	\subfigure{\includegraphics[width=0.83\columnwidth]{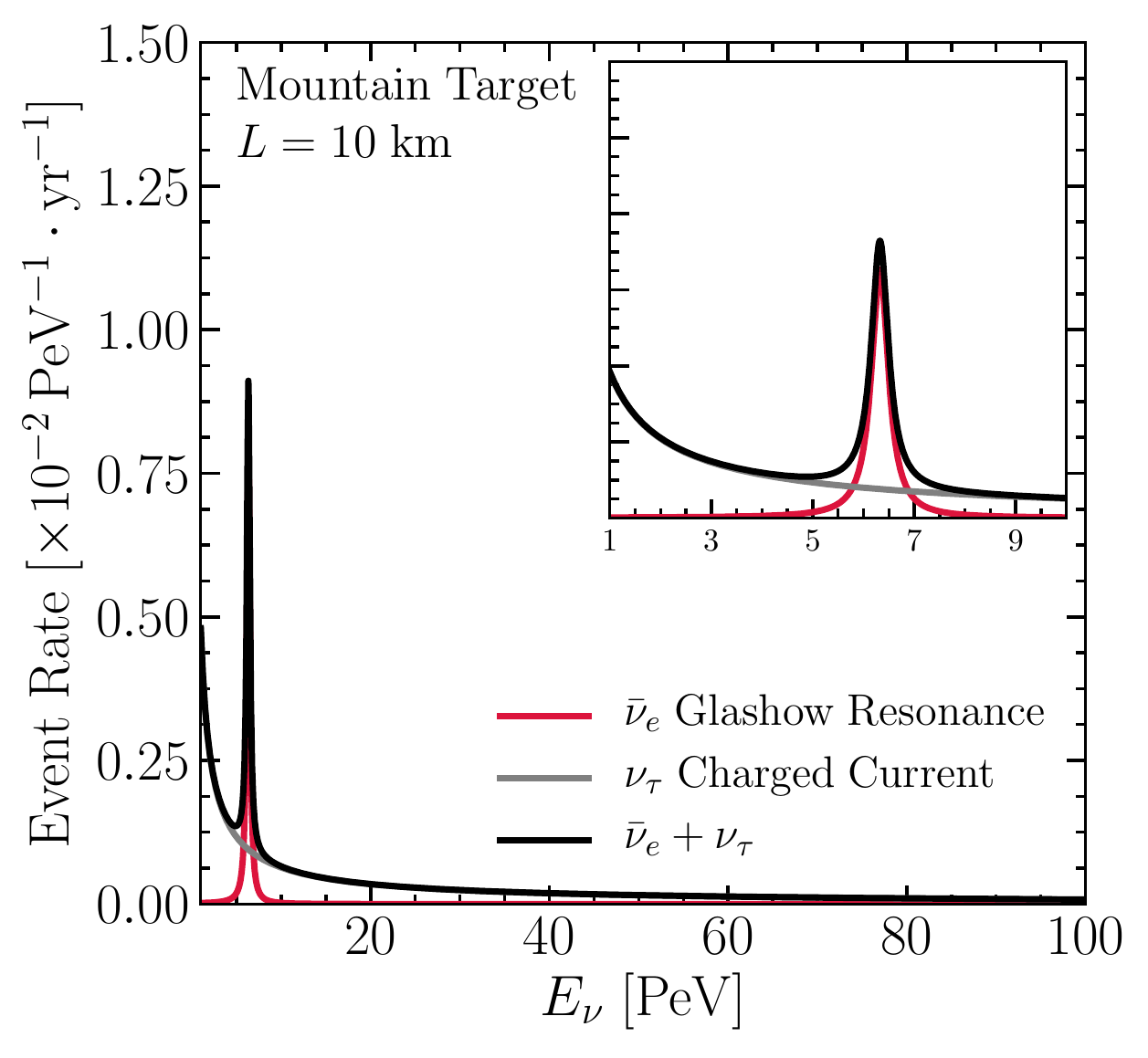}} \hspace{1cm}
	\subfigure{\includegraphics[width=0.83\columnwidth]{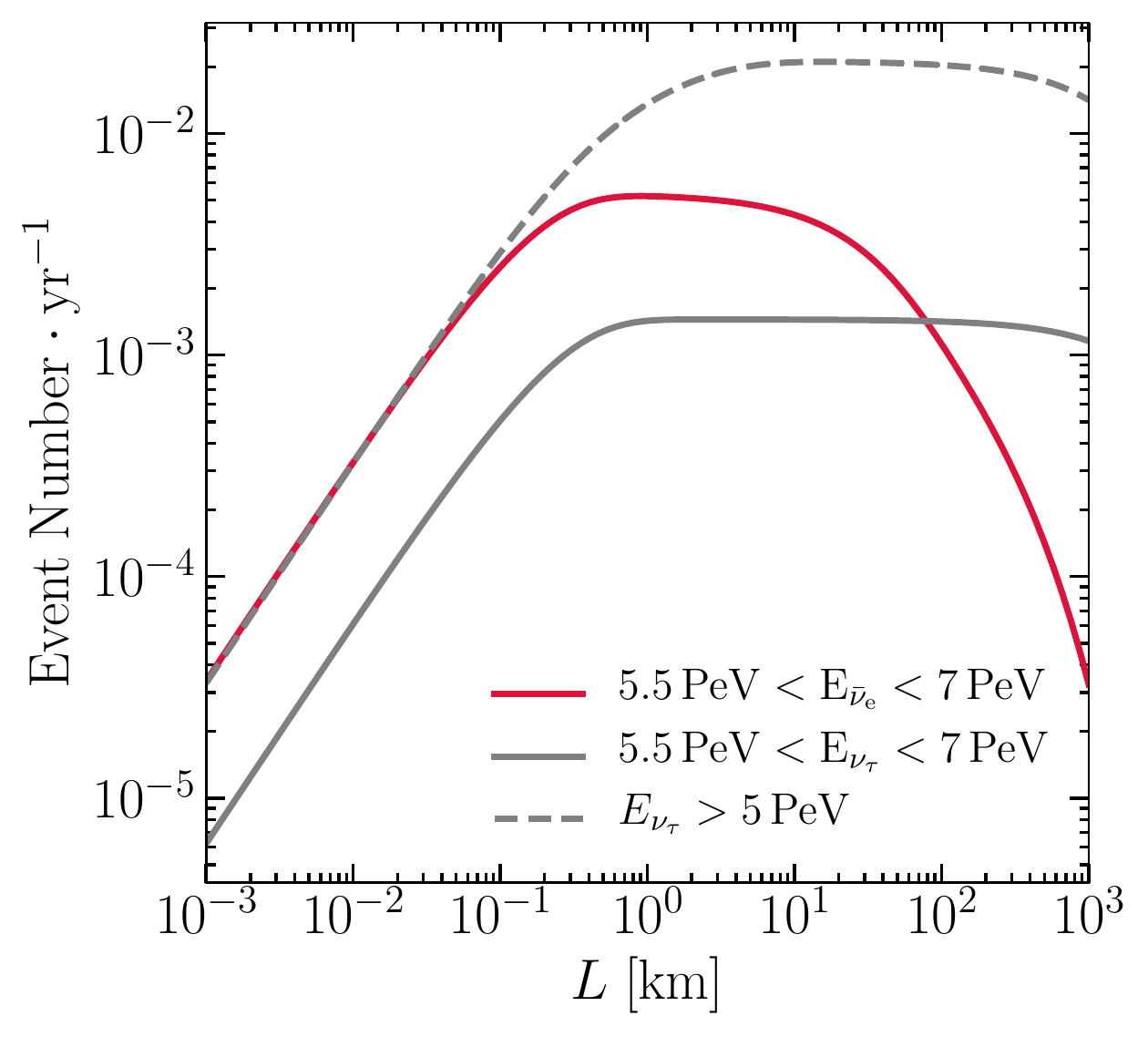}}
	\caption{\emph{Left-panel}: the event rate per neutrino energy of mountain-penetrating neutrinos with an effective tau aperture $A^{\tau}_{\rm geo} \simeq 1~{\rm km^2 \cdot sr}$. The mountain thickness has been fixed as $10~{\rm km}$. \emph{Right-panel}: the event number per year within the indicated energy range as a function of the mountain thickness.\label{fig:MTEV}}
\end{figure*}

In Fig.~\ref{fig:MTCE} we display the conversion efficiencies of both the Glashow channel and the CCDIS case as a function of the neutrino energy. A notable excess around the resonance can be observed for neutrino energies from 5.5 PeV to 7 PeV. The band is generated by varying the mountain thickness from 0.1 km to 50 km. The efficiency peak corresponding to the Glashow resonance is almost one order of magnitude larger than the CC one.

Not all emerged taus from the mountain surface can be detected by the Cherenkov light collector. For the case of the diffuse astrophysical neutrinos, a geometric aperture can be defined to describe the acceptance that an escaped tau be observed: $A^{\tau}_{\rm geo} \simeq S \cdot \Delta \Omega$, with $S$ being the surface area of the mountain facing the detector, and $\Delta \Omega$ being the solid angle covered by the Cherenkov cone. 
This is a good approximation due to the fact that taus at PeV energies will decay immediately after they escape from the mountain.
The mountain surface area can be typically set as $S \simeq 100~{\rm km^2}$. 
The solid angle covered by the Cherenkov emission is determined by several factors, e.g., the tau energy prior to its hadronic decay, the distance from the shower location as well as the photon detection capability of the sensors. 
A value of $\Delta \Omega \simeq 0.024~{\rm sr}$ might be taken, corresponding to a Cherenkov angle of $5^{\circ}$.
For simplicity, we fix the geometric aperture for the tau detection of the nominal telescope as $A^{\tau}_{\rm geo} \simeq 1~{\rm km^2\cdot sr}$. The final event rate can be obtained with 
$
{\mathrm{d}N^{}_{\rm mt}}/({\mathrm{d}E_{\nu} \mathrm{d}T}) = \Phi^{}_{{\nu}} \, A^{\tau}_{\rm geo} \, \mathcal{P}^{}_{\nu \rightarrow \tau}.
\label{eq:eventMountain}
$

We display the differential event rates in the left panel of Fig.~\ref{fig:MTEV}. 
The best-fit diffuse flux given by the IceCube TMG measurement has been extrapolated to $100~{\rm PeV}$ and is given as the input.
A clear bump due to the Glashow resonance of $\overline{\nu}^{}_{e}$ arises above the background of $\nu^{}_{\tau}$ events. The event ratio of $\overline{\nu}^{}_{e}$ to ${\nu}^{}_{\tau}$ in the region of the bump ($5.5~{\rm PeV}$ to $7~{\rm PeV}$) is as large as 3. 
In the right panel of Fig.~\ref{fig:MTEV} we compare the $\overline{\nu}^{}_{e}$ event number inside the resonant bump to the $\nu^{}_{\tau}$ one as a function of $L$. For a mountain thickness from $1~{\rm km}$ to $100~{\rm km}$, the ratio of $\overline{\nu}^{}_{e}$ events to $\nu^{}_{\tau}$ background in the range of $5.5$-$7~{\rm PeV}$ ($5$-$100~{\rm PeV}$) changes from $3.7$ to $0.8$ (from $0.4$ to $0.05$). 
For very thin mountains, the target volume is restricted, while for very thick ones, the $\overline{\nu}^{}_{e}$ near the resonance would be attenuated. An ideal choice of the mountain thickness for the $\overline{\nu}^{}_{e}$ detection is $1$-$10~{\rm km}$.

However, with the acceptance of $A^{\tau}_{\rm geo} \simeq 1~{\rm km^2\cdot sr}$, the integrated event number for both $\nu^{}_{\tau}$ and $\overline{\nu}^{}_{e}$ are too low for the diffuse neutrino flux measured by IceCube, e.g., $N^{}_{\overline{\nu}^{}_{e}} \simeq 0.005~{\rm yr^{-1}}$ and $N^{}_{\nu^{}_{\tau}} \simeq 0.02~{\rm yr^{-1}}$ for $E^{}_{\nu} > 5~{\rm PeV}$ with the TGM best-fit flux. The event rate in a realistic experiment is subject to the actual profile of the mountain and the detector configuration. One way to improve the sensitivity might be the deployment of array of Cherenkov detectors, such that a larger total solid angle of the diffuse flux can be accessed. The fluorescence detector can also be utilized to increase the angular acceptance.
The sensitivity to neutrinos from a specific point source behind the mountain could be higher. In a similar way, the acceptance of the parallel tau flux induced by neutrinos from the point source can be described by the geometric area $A^{\tau}_{\rm geo} \simeq  D^2/2 \cdot \Delta \Omega$, with $D$ being the distance from the Cherenkov detector to the mountain surface. Given $D = 25~{\rm km}$, the effective area is estimated to be $7~{\rm km^2}$, however, the actual number may vary by orders of magnitude depending on the experimental details.

Here we take the planned Neutrino Telescope Array (NTA) project as an example, as its sensitivities to both point source and diffuse flux are comparable to or even better than those of IceCube in the PeV to EeV energy range. NTA can achieve a very competitive sensitivity by observing the Earth-skimming and mountain-penetrating neutrinos. Both Cherenkov light and fluorescence detectors will be installed. For the case of the diffuse flux, one may infer the geometric area for tau acceptance from Fig.~17 of Ref.~\cite{Sasaki:2014mwa}. Taking the conversion efficiency of $\nu^{}_{\tau}$ at 6.3 PeV with a mountain width of $10~{\rm km}$ to be $5\times 10^{-5}$, the geometric aperture for the tau from the diffuse neutrino flux may be speculated as $60~{\rm km^2\cdot sr}$, much larger than our previous toy setup. 
The enhancement of the acceptance should be attributed to the multi-site coverage as well as the deployment of the fluorescence detector.
With this experimental input, we find the total Glashow event rate is $0.3~{\rm yr^{-1}}$ with the best-fit spectrum of IceCube TGM sample. The p-value
to observe at least one neutrino event for one year full exposure around the Glashow peak is $26\%$. 
If we adopt the softer HESE best-fit spectrum instead, the rate will decrease to $0.03$ per year. 
Some earth-skimming neutrinos should also have contributed to the sensitivity of NTA for diffuse flux, which will decrease the Glashow rate estimation here.
So we conclude that it is viable but a little challenging to observe the Glashow resonance signature from the diffuse neutrino flux in the mountain-penetrating type telescopes of the near future. 
A dedicated Monte-Carlo simulation is necessary to make a more solid conclusion.
On the other hand, a positive detection of such event is a good complement to the IceCube cascade event searches \cite{Lu:2017nti,lulu}, of great scientific importance to the neutrino astronomy.

\section{Concluding remarks}
The Glashow resonant scattering enhances the neutrino cross section in matter by orders of magnitude with respect to the CCDIS case.
For the cases of through-going events at IceCube and the detection of Earth-skimming neutrinos, the Glashow resonance will lead to an event excess higher than $10\%$ near the horizontal direction of the telescope. 
The expected event number at IceCube with ten years of data taking is only $0.28$ within the $10\%$-contour. A similar conclusion holds for the detection of Earth-skimming neutrinos, e.g. only 0.024 events per year within the $10\%$-contour is expected for a rather optimistic proposal like CHANT. For IceCube, the cascades search is still the best window to observe the Glashow resonance events.

In contrast, the detection of mountain-penetrating neutrinos is potentially sensitive to the Glashow resonance signature, as the solid angle will not shrink at resonance as in the former two cases. However, a low detection threshold down to 6.3 PeV must be achieved in order to efficiently probe $\overline{\nu}^{}_{e}$ via the Glashow resonance channel.
The Cherenkov light detection complemented by the fluorescence technique, e.g. in NTA, is promising for $\overline{\nu}^{}_{e}$ Glashow resonance detection if with enough exposure and detection efficiencies.

Our study here can be straightforwardly extended to the searches of new physics \cite{Jezo:2014kla,Babu:2019vff}, such as the type-II seesaw and non-standard interactions. The resonant scattering mediated by heavy new particles in those PeV-EeV telescopes can be a promising channel to probe new physics beyond the Standard Model. 
We will leave this interesting possibility for a future study.
\\
\begin{acknowledgments}
\sl The authors would like to thank Carlos Arg{\"u}elles, Francis Halzen and Shun Zhou for useful discussions and comments on the manuscript. GYH is supported in part by the National Natural Science Foundation of China under grant No.~11775232. QL is supported by NSF under grants PLR-1600823
and PHY-1607644 and by the University of Wisconsin Research Council with funds granted
by the Wisconsin Alumni Research Foundation.
\end{acknowledgments}

%%%%%%%%%%%%%%%%%%%%%%%%%%%%%%%%%%%%%%%%%%%%%%%References%%%%%%%%%%%%%%%

%\bibliographystyle{plainnat}
\bibliography{references}

\clearpage

\end{document}